\documentclass[aps,twocolumn,pra,showpacs,floatfix]{revtex4}
\usepackage[dvips]{graphicx}
\usepackage[usenames]{color}
\usepackage{dcolumn}
\usepackage{bm}

\def\be{\begin{equation}}
\def\ee{\end{equation}}
\def\bea{\begin{eqnarray}}
\def\eea{\end{eqnarray}}

\begin{document}

\author{Jakub S. Prauzner-Bechcicki$^1$, Krzysztof Sacha$^1$, 
Bruno Eckhardt$^2$, and Jakub Zakrzewski$^1$}
\affiliation{
$^1$Instytut Fizyki Mariana Smoluchowskiego and
Mark Kac Complex Systems Research Center \\
  Uniwersytet Jagiello\'nski, Reymonta 4, 30-059 Krak\'ow, Poland \\
$^2$Fachbereich Physik, Philipps-Universit\" at Marburg, D-35032
Marburg, Germany}

\title{Quantum model for double ionization of atoms in strong 
laser fields}

\date{\today}

\begin{abstract}
We discuss double ionization of atoms in strong laser pulses using a reduced
dimensionality model. Following the insights obtained from an analysis of the
classical mechanics of the process, we confine each electron to move along the lines
that point towards the two-particle Stark saddle in the presence of a field.
The resulting effective two dimensional model is similar to the aligned electron model,
but it enables correlated escape of electrons with equal momenta, as observed 
 experimentally.
The time-dependent solution of the Schr\"odinger equation allows us to discuss in detail 
the time dynamics of the ionization process, the formation of electronic wave packets and 
the development of the momentum distribution of the outgoing electrons. 
In particular, we are able to identify the rescattering process, simultaneous direct
double ionization during the same field cycle, as well as other double ionization processes.
We also use the model to study the phase dependence of the ionization process.
\end{abstract}
\pacs{32.80.Rm,32.80.Fb,03.65.-w,02.60.Cb}%,02.60.-x}

\maketitle

\section{Introduction}

Extensive experimental and theoretical studies of multiple ionization in 
strong laser fields have revealed a number of unexpected features. 
While some phenomena can be described as multi-photon, independent electron
processes, (see e.g. \cite{gelt88}), others require electron-electron
correlations. This was concluded on the basis of a detailed analysis of 
experimental data together with precise calculations of single ionization rates 
\cite{anne82,review}. Several experiments \cite{walker94} then revealed a pronounced 
``knee'' structure in the yield vs peak laser
intensity, typically plotted with logarithmic axis because of the wide range
of values covered. Further ingenious experiments that resolved the
joint momentum distributions of the outgoing electrons then showed that 
they often leave the atom with the same momenta \cite{review,weber00n}. 
This has triggered a number of theoretical studies of this process, including
$S$-matrix calculations for the full cross sections \cite{becker00kopold00}, and
investigations of simplified classical and quantum models. Among the models 
are so-called aligned-electron models \cite{aligned,engel}, in which electrons move in 
a one-dimensional (1D) regularized Coulomb potential, or quasi three-dimensional (3D) ones with
the center of mass of the electrons confined to move along 
the field polarization axis \cite{becker}. On the other hand, 
an exact solution of the time-dependent 
Schr\"odinger equation for two electrons in a laser field remains a formidable task,
accessible usually to very short pulses of wavelengths shorter than 800~nm
\cite{taylor,parker06}.

One of the keys to understanding the dynamics of double and higher multiple-ionization 
is the rescattering scenario
\cite{corkum93}. 
%While most electrons 
%that escape from the atom when the field is strong
%escape and contribute to the single ionization channel, some have their paths 
Most of the electrons
 escaping from the atom when the field is strong, leave definitely
and contribute to the single ionization channel.
Some, however,  have their paths reversed back to the ionic core when the field changes sign. These electrons are then accelerated by 
the field and can share their energy with one or more electrons close to the nucleus. 
When they return to the ion, the electrons form a transient, short lived compound state 
which has several possible channels to decay:
it can exit in a single ionization event, a double ionization event or a repetition of the rescattering
cycle. Interestingly, starting from this intermediate situation,  a classical analysis easily yields 
possible pathways to ionization and the effective potential \cite{eckhardt01pra1}. 
The classical analysis of double ionization suggests that 
the electrons may escape simultaneously if they pass sufficiently 
symmetrically over saddles that form in the presence of the electric field. 
As the field phase changes, the saddles move along lines that keep a constant angle 
with respect to the polarization axis \cite{eckhardt01pra1}. 

The observation about the location of the saddles then suggested a way to 
reduce the degrees of freedom but to allow the possibility for the two electrons
to escape with the same momentum\cite{eckhardt06}: as in the aligned electron models,
the electrons are confined to move along lines, except that the lines pass through the
location of the saddles and have an angle of 
$\pi/6$ with respect to the field axis.
As other  1D+1D models, also this one allows to reproduce \cite{prauzner07} tunnelling and 
rescattering processes,  single and sequential double ionizations but, in addition, it 
allows for escape with the same momentum and hence can mimic the correlated electron 
escape. In the aligned--electron model this process is suppressed by the overestimated 
Coulomb repulsion.

The aim of the present paper is to present the quantum version of this model
and to discuss in detail its predictions for the ionization signal. Results for longer pulses
have been described briefly before \cite{prauzner07}. The model is introduced and motivated
in section \ref{physics}. Section \ref{methods} 
then contains technical information on numerical aspects of the 
calculations. Except for section \ref{config_space} 
on the methods used to distinguish the different 
ionization signals much of the material can be skipped on a first reading.
In section \ref{pulses} we focus on the results for short pulses and the temporal sequence of events.
In section \ref{fringes} we focus on the modulations which appear in the momentum distributions and
which we related to the rescattering mechanism. We conclude with some final remarks in 
section \ref{end}.

\section{Physical foundations of the simplified dynamics}
\label{physics}

Consider a Hamiltonian for a non-relativistic He atom (in atomic units),
\begin{equation}\label{hamhe}
H=\sum_{i=1}^2\left(\frac{{\bf p}_i^2}{2}-\frac{2}{|{\bf r}_i|}\right)+
\frac{1}{\sqrt{({\bf r}_1-{\bf r}_2)^2}}+H_{int},
\end{equation}
where $H_{int}$ is the part of Hamiltonian describing the interaction with the 
external field. Its detailed form depends on the gauge, and different forms have
different advantages. For the calculation of the ionization yields, 
we will use the position gauge. Then, 
for a laser pulse linearly polarized along the $z$ axis, $H_{int}$ takes the form:
\begin{equation}
H_{int} = F(t)(z_1+z_2),
\end{equation}
where the electric field is composed as
\be F(t) = F _0 f(t) \sin(\omega t +\phi),
\label{pole}
\ee
with $F_0$, $f(t)$, $\omega$ 
and $\phi$ being the peak amplitude, the envelope, the frequency and 
the initial phase, respectively.  In the following we 
assume  the frequency corresponding to the wavelength of 800~nm,
i.e. $\omega=0.06$~a.u., and the sine-squared envelope,
\be
f(t)=\sin^2\left(\pi t/T_d\right),
\label{envelope}
\ee
where $T_d$ is the pulse duration.

The velocity gauge is more convenient for evaluation of
momenta distributions: then  $H_{int}$ reads
\begin{equation}
H_{int} = A(t)(p_{z1}+p_{z2})+A(t)^2/2,
\end{equation}
with the vector potential $A(t) = - \int_0^t F(t') \rm{d}t' $.

Solution of the Schr\"odinger equation corresponding to (\ref{hamhe}) 
is a formidable numerical task
\cite{taylor,parker06} involving six spatial dimensions. For visible or infrared 
frequencies it requires supercomputer resources. Similar restrictions 
apply to S-matrix based calculations \cite{becker00kopold00}. Thus
simplified models that allow for reduction of the dimensionality of the problem
are desirable. For single electron
ionization under the influence of linearly polarized wave a
1D model where the electron is restricted to move along the 
field polarization axis was proposed already more than 20 years ago \cite{jensen84}.
Such a model was shown to capture the essence of the ionization process both in 
microwave \cite{leopold89} as well as the optical \cite{eberylium} domain. 
A similar 1D reduction was soon applied to two electron atom \cite{aligned,engel}.
The technical advantages of such a 1D+1D electron model are numerous and 
the reduction appeared justified as the model seemed to 
capture essential features of experiments. 
However, the model had to be called into doubt 
with the advent of experiments that showed that  a significant
fraction of electrons leaves the atom simultaneously, in a correlated manner, 
with the same momenta parallel to the polarization axis. Clearly, the collinear
1D+1D models overestimate the effects of the Coulomb repulsion, and while they
can model processes in which the electrons ionize at vastly different times, they
cannot capture such a simultaneous ionization event.

To overcome this shortcoming of the aligned electrons model a different
reduction has been proposed in \cite{becker}: the motion of the center of 
mass of the system was restricted to move along the 
polarization axis. This reduces the effective dimensionality 
of the problem from six to three spatial dimensions, and enables 
numerical simulations. The obvious drawback of the model is that it 
introduces unusual long-range correlations between the electrons which can be 
expected to be small when both electrons are at about the same distance 
from the 
nucleus, but which may change the dynamics when one electron is close to the nucleus  and the other far away.

\begin{figure}[t]
\begin{center}
\includegraphics[width=0.4\textwidth,clip]{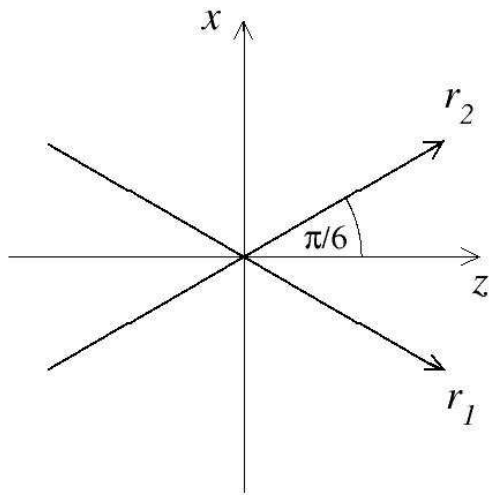}
\end{center}
\caption[]{Lines in configuration space along which the electrons are allowed to move. $r_i$ denote the
position of the electrons along the lines. The angle of $\pi/6$ is chosen such they agree with 
the motion of the Stark saddle under variation of the field.}
\label{geo1}
\end{figure}

\begin{figure}[t]
\begin{center}
\includegraphics[width=0.4\textwidth,clip]{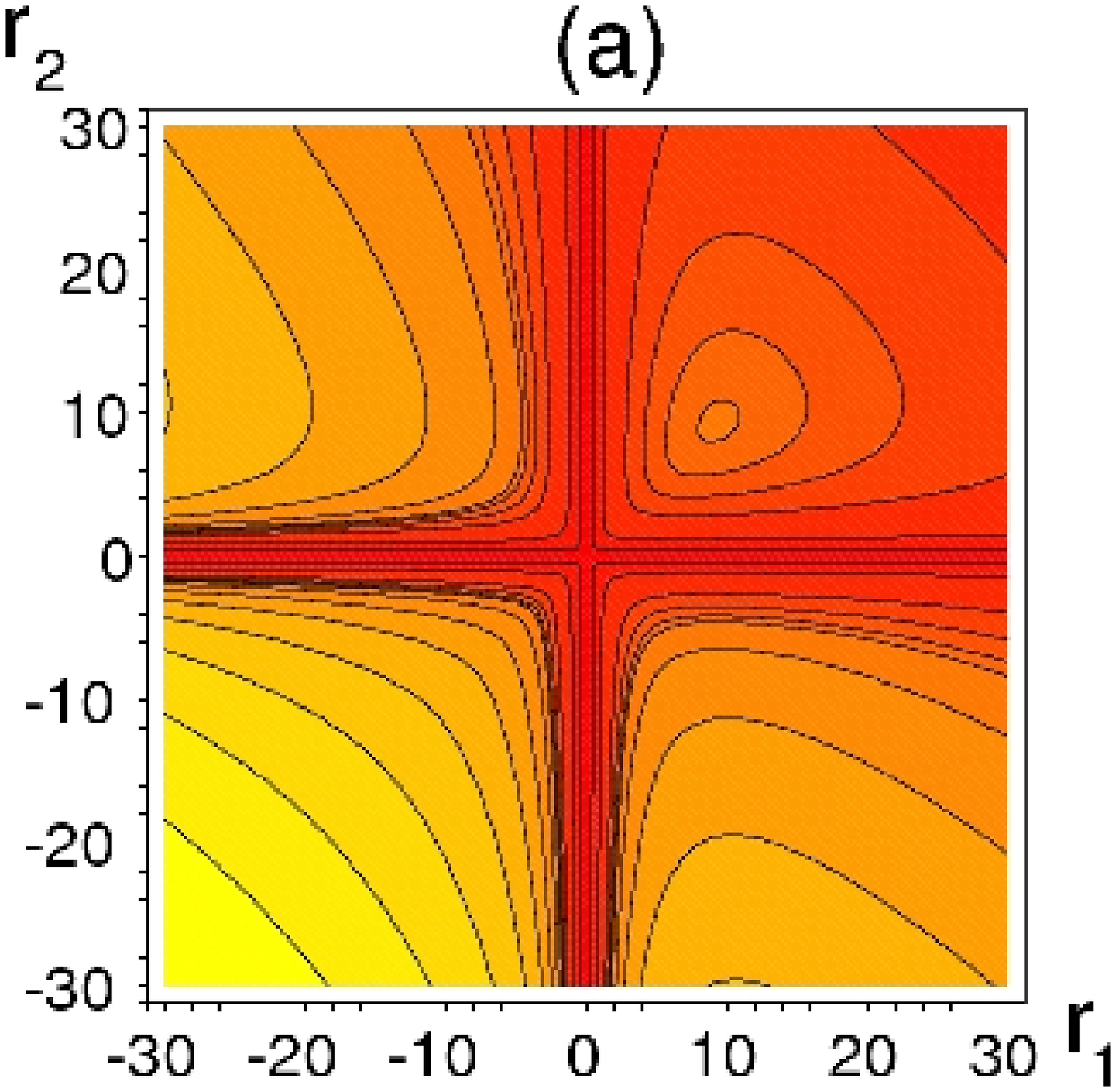}
\includegraphics[width=0.4\textwidth,clip]{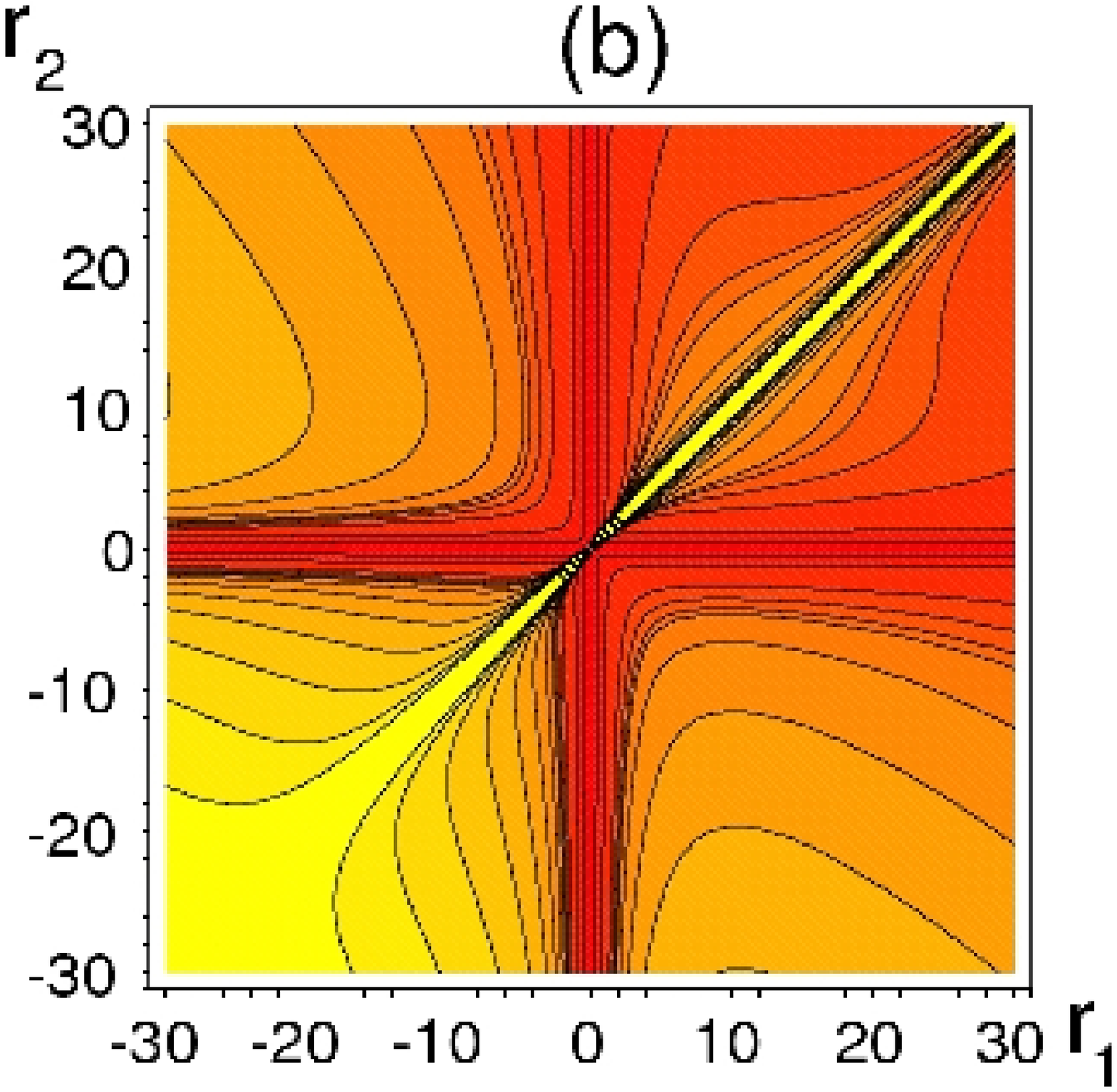}
\end{center}
\caption[]{(Color online) Comparison between the potentials (a) for the Hamiltonian (\ref{ham}) used here and
(b) the aligned electron model. In the latter case both electrons are restricted to move along the 
polarization axis, and the diagonal is blocked by the diverging Coulomb repulsion.
}
\label{poten}
\end{figure}

A model that keeps the simplicity of the aligned electron models but allows for correlated
electron escape was proposed in \cite{eckhardt06}. The model builds on the insights gained
from classical trajectory studies \cite{eckhardt01pra1}. In the presence of the field, each
electron sees an effective Stark saddle.
Due to electron-electron repulsion
the saddles move from the repulsion free positions and are placed symmetrically
with respect to the polarization axis and  
within the same distance from the nucleus. 
%The separation between the electrons increases with distance, since 
%the attraction to the nucleus decreases. 
%As both attraction and repulsion scale in the same way, 
The saddles move along straight lines  when the field changes. 
The observation
that the electrons have to cross this saddle then led to the suggestion to restrict the 
electronic motion to the lines traced by the saddles.
The lines form angles of $\pm \pi/6$ with respect to the 
polarization axis (compare Fig.~\ref{geo1}). Since electrons 
separate as they simultaneously move out from the nucleus their mutual repulsion
 diminishes, and the resulting potential far from the diagonal, i.e. around  $r_1=r_2$ (where $r_1$ and $r_2$ are the positions along 
the saddle lines),
becomes quite similar
to that of the aligned electron model, as evident from 
the potential landscapes in Fig.~\ref{poten}. 
The two potentials differ near the diagonal,
where both electrons have the same distance from the nucleus: the diagonal is accessible
in the present model, but blocked by Coulomb repulsion in the aligned electron 
model. 

As discussed in \cite{prauzner07}, this 1D+1D restricted model is able to
reproduce the tunnelling and the rescattering processes, as well as a subsequent
simultaneous double ionization or other paths to double and single 
ionization. From the classical 
mechanics point of view the model has the drawback that the
chosen subspace is not an invariant subspace of the full motion, as is the case for 
the aligned electron model. This drawback is shared with the previously mentioned 
model of constrained center of mass motion \cite{becker}. 
Nevertheless, we will see that the model does yield qualitative predictions and
valuable insights into the relevant processes and interactions.

In the new ``saddle-track'' coordinates the dynamics of two electrons in the linearly 
polarized laser field is given by the Hamiltonian \cite{eckhardt06}:
\begin{equation}\label{ham}
H=\sum_{i=1}^2\left(\frac{p_i^2}{2}-\frac{2}{|r_i|}+\frac{F(t)\sqrt{3}}{2}r_i\right)+
\frac{1}{\sqrt{(r_1-r_2)^2+r_1r_2}},
\end{equation}
where $r_1$ and $r_2$ are the electron coordinates along the saddles' lines. 

\section{Methods and Procedures}
\label{methods}

This section is mainly technical and devoted to details of the numerical
procedures used in the following sections. 
The subjects
covered in this section include the numerical method in \ref{technical_numerics},
the partitioning of configuration space in \ref{config_space}, the
calculation of the ionization yields in \ref{technical_ionisation} and
the extraction of the final momentum distributions in \ref{technical_final}.
Readers interested mainly in 
the physical results should read the description in section \ref{config_space}
of how the different ionization channels are identified in the configuration space, 
and may then proceed directly to the next section. 

\subsection{Numerical methods}
\label{technical_numerics}
The simplified 1D+1D model is applied to calculate ionization yields 
for single and 
double ionization as well as to obtain electron and ion momenta distributions.  
The Schr\"odinger equation corresponding to the Hamiltonian (\ref{ham}) is 
solved on a grid using the operator splitting method combined with the 
Fast Fourier Transforms to effectively switch between the position (appropriate 
for the potential) and the momentum (for the kinetic energy evaluation) representations. 
The ionization yields are efficiently obtained in the length gauge while the velocity 
gauge is used for the momenta distributions. In the latter case, the physical space 
is divided into different regions with Coulomb repulsion neglected in the outer regions 
(as explained in details below). 

The potential singularities in (\ref{ham}) are removed by replacing $1/x$ by 
$1/\sqrt{x^2+e}$ with $e=0.6$. 
This leads to a ground state energy of the unperturbed atom of $E_g=-2.83$ 
(calculated by means of the imaginary time evolution).

In the time evolution, in order to minimize the undesired reflections at 
the edges of the integration region, absorbing boundary conditions are included 
by adding imaginary potentials 
\be
V_{j}=\left\{
\begin{array}{c}
 - i \eta (|r_j|-x_0)^{\alpha}, \quad {\rm for} \quad |r_j|>x_0, \\
 0, \quad {\rm elsewhere}, 
\end{array}\right.
\ee
where $x_0$ is the distance from the center (along each axis) beyond which
 the imaginary potential becomes active. 
The value of $x_0$ is chosen sufficiently large so as to not perturb the dynamics close 
to the nucleus, but it is smaller than $L/2$, 
when the integration domain in one direction is $[-L/2,L/2]$.
The parameters $\eta$ and $\alpha$ are optimized with respect to the distance 
from the edge of the grid, $L/2-x_0$, on which absorbing boundary conditions 
are implemented. We use $\eta=10^{-5}$ and $\alpha = 4$                                         .

\subsection{Identifying outgoing channels}
\label{config_space}

\begin{figure}
\includegraphics[width=0.4\textwidth,clip]{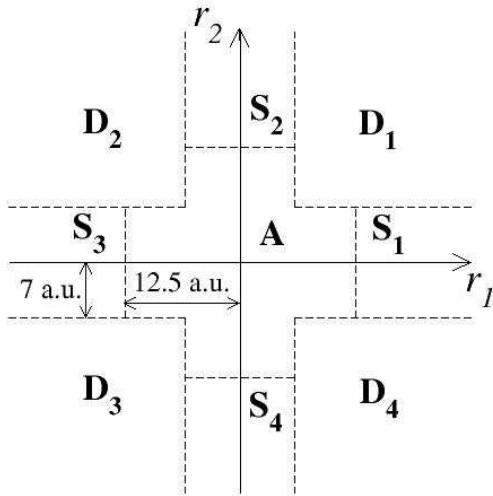}
\caption[]{Configuration space of the model. The regions labelled 
{\bf A}, {\bf S}$_i$ and {\bf D}$_i$  
correspond to the neutral atom, and singly charged ion, and doubly charged ion populations, 
respectively. Ionization rates are estimated from the flux across the appropriate 
borders, and momentum distributions are obtained by propagating the fractions of the wave
functions in the corresponding singly and doubly ionized regions
(see section \ref{config_space} for more details).
\label{geometry}}
\end{figure}

For the proper assignment of the final state to the appropriate decay channel 
we have to identify indicators for single and double ionization.
Some guidance as to how to pattern configuration space is provided by the interaction potentials
and the regions in which they dominate. As in many other cases, the long range nature of the
Coulomb interaction calls for special care, but we will here adopt a pragmatic approach and simply
classify state space by the regions in which the interaction is stronger than some threshold.
This then results in the division of state space as shown in Fig.~\ref{geometry}. 
We follow in this respect the original ideas developed in the Belfast group \cite{taylor}.
When both electrons are close to the core and interact strongly, the system is in an atomic state,
so this region is labelled ${\bf A}$. 
If one electron escapes along the $r_1$-axis and the other is trapped, i.e. $r_2$ is bounded,
then only the attraction of this second electron to the nucleus remains asymptotically: 
this defines the bands parallel to $r_1$, which are labelled ${\bf S}_1$ and ${\bf S}_3$ 
to indicate single ionization (in this case, of electron $1$). 
Similar considerations lead to the definition of ${\bf S}_2$  and ${\bf S}_4$ 
for the single ionization of electron $2$. If both electrons
escape, then we have double ionization, indicated by the four regions ${\bf D}_i$. 
The numerical values for the borders between different regions affect the
quantitative predictions of the model to some extend. This is not a major
problem here since anyway the restricted dimensionality model we consider yields
qualitative predictions only. We take the original Belfast proposition \cite{taylor} for numerical values entering the model (indicated in 
Fig.~\ref{geometry}).

Starting from a state localized near the nucleus, there are several paths to
double ionization. A sufficiently strong field
will open a Stark saddle along the diagonal, thus enabling the electrons to 
pass directly from ${\bf A}$ to ${\bf D}_1$ and ${\bf D}_3$, respectively. We call
this simultaneous escape (SE), since both electrons move into the ionized region
directly. The processes called REDI for 
{\bf RE}collision induced {\bf D}irect {\bf I}onization  \cite{weber00n} 
are one example of this behaviour. 
Other contributions could come from situations in which the
doubly excited state formed during rescattering persists for a while
and does not decay until a later field cycle.

We can also link the different quadrants of our model with the Z and NZ trajectories discussed  by Ho et al\cite{Ho2005}. The Z trajectories have zero 
or small momenta
for the ion, and thus require the electrons to move in opposite directions.
A Z trajectory will hence end up in section ${\bf D}_2$ or ${\bf D}_4$. 
Such events are allowed, but, according to our calculations,
very rare. An NZ trajectory, where the momentum of the ion is non-zero,
will end up in sector ${\bf D}_1$ or ${\bf D}_3$.
Note that only the NZ trajectories are influenced by the electron repulsion 
during their escape, and hence only they will show the momentum correlations 
in the final state (according to the classical analysis in \cite{eckhardt01pra1})

Other paths to double ionization pass through the single ionization regions ${\bf S}_i$ before
entering one of the ${\bf D}_j$. Such paths are then either purely
independent sequential electrons escapes or  reminiscent of the
RESI processes in longer pulses: {\bf R}ecollision and {\bf E}xcitation of the 
ion plus {\bf S}ubsequent {\bf I}onization \cite{RESI}.
We will collect their signal under the common name: a 
consecutive escape (CE).

\subsection{Ionization yield}
\label{technical_ionisation}

With the outgoing channels properly identified, 
the determination of the physical observables becomes relatively easy. For instance,
population of one of the states can be determined by integrating the modulus squared 
of the wavefunctionover a given region. E.g. the population of atoms follows
from 
\be 
P_A = \int_A |\psi|^2 {\rm d}r_1{\rm d}r_2. 
\ee
However, due to the fact that absorbing boundary conditions are applied, the 
wavefunctionleaks out of the interaction region, and the
norm of the total wavefunctionand the calculated yields decrease with time.
To overcome this problem, the yields are instead calculated by 
using the probability fluxes between appropriate regions, 
as suggested by Dundas {\it et al.} \cite{taylor}. 
The fluxes are obtained from the quantum mechanical continuity equation,
\begin{equation}
\label{cont_eq}
\frac{\partial}{\partial t}\rho + {\bf \nabla}\cdot{\bf j} = 0,
\end{equation}
where $\rho = |\psi|^2$ is the probability density 
and ${\bf j} = -i(\psi^{\star}{\bf \nabla}\psi - \psi {\bf \nabla}\psi^{\star})/2$ 
is the probability current \cite{footnote1}.
Integration of the continuity equation over  a given region 
($R \in \{ {\rm {\bf A}, {\bf S}, {\bf D}} \}$)  and over time
gives the population of the corresponding species:
\begin{equation}
 P_R(t) = - \int\left( \int_R {\bf \nabla}\cdot{\bf j} {\rm d}r_1 {\rm d}r_2
\right) {\rm d}t 
= - \int \mathcal{F}_R(t) {\rm d}t,
\end{equation}
where $\mathcal{F}_R(t)$ is the probability flux over boundaries of the region $R$.

\begin{figure}[t]
\vspace{0.7cm}
\begin{center}
\includegraphics[width=0.4\textwidth,clip]{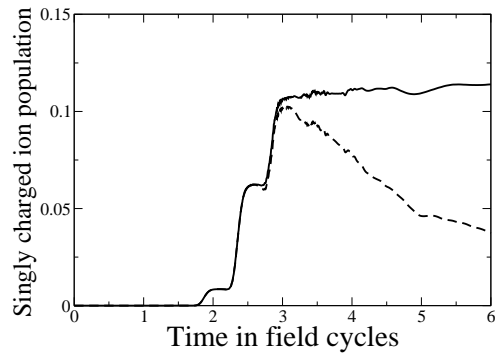}
\end{center}
\caption[]{Comparision between the singly charged ion population as calculated using different methods: 
direct integration of the modulus squared wavefunctionover the regions {\bf S$_i$} (dashed line) and 
integration of the probability fluxes through the boundaries of the {\bf S$_i$} regions (solid line). 
Absorbing boundary conditions were introduced at a distance 150 a.u. from the nucleus and a 5 cycle laser 
pulse with amplitude $F_0=0.19$ a.u. was used.}
\label{sin}
\end{figure}

Fig.~\ref{sin} compares the population of singly charged ions as 
a function of time
obtained applying both methods described above. The results are 
for a 5 cycles long laser pulse with amplitude  $F_0=0.19$. 
To allow the ionised (possibly slow) electrons to leave the vicinity of the nucleus, 
here and in all the other calculations the evolution is continued for one additional cycle. 
The absorbing boundary conditions are set 
at the distance of $x_0=150$~a.u. from the nucleus. As can be seen, as long as the electrons do 
not reach the absorbing boundaries both methods give the same result. Then, once electrons 
are absorbed, there is a dramatic fall of the population calculated
from the direct integration of the modulus squared wavefunctionover
the {\bf S$_i$} regions. However, the ionization yield calculated from the probability fluxes 
remains essentially flat.

To obtain the double ionization yield we calculate fluxes through the boundaries 
of the {\bf D$_i$} regions. This in turn enables us to distinguish between
the direct and indirect double ionization events by calculating probability 
fluxes over relevant boundaries, as discussed before in section \ref{config_space}.
That is, one calculates the flux between regions corresponding to singly and doubly charged ions
in order to get the RESI ionization yield, and between regions {\bf A} 
and {\bf D$_i$} to get the direct double ionization yield. 

To summarize:
\begin{itemize}
\item The population of single ions (SI) at time $t$ is obtained 
from time integration of the fluxes from {\bf A} to {\bf S$_i$} ($i=1,2,3,4$) 
minus the fluxes from {\bf S$_i$} to {\bf D$_j$};
\item The probabilities for simultaneous double escape (SE) are 
obtained from the time integration
of the fluxes {\bf A}$\rightarrow${\bf D$_1$} and 
{\bf A}$\rightarrow${\bf D$_3$}.
\item A possible simultaneous, anti-correlated double ionization probability is obtained 
from the time integrals of the fluxes {\bf A}$\rightarrow${\bf D$_2$} and 
{\bf A}$\rightarrow${\bf D$_4$}.
\item Integration of fluxes from {\bf S$_i$} to {\bf D$_j$} ($i,j=1,2,3,4$)
then gives a measure of  
the consecutive escape (CE), i.e.  sequential double
ionization and RESI type processes.
\end{itemize}

\subsection{Momentum distributions}
\label{technical_final}

At first glance it seems possible to obtain a distribution of electron momenta that corresponds 
to double ionization by calculating the Fourier transform of the part of the 
wavefunctionlocalized in the {\bf D$_i$} regions (see Fig.~\ref{geometry}), using the fact
the squared modulus of the wavefunctionin the momentum representation gives a momentum distribution. 
However, since absorbing boundary conditions are applied, the information contained
in the final wavefunctionis not complete: the faster electrons  reach  quickly the edges of the domain where they are absorbed, and the information about their momenta is lost. 

To overcome this problem, we use a method proposed by Lein et al.~\cite{engel}.
It is based on the observation that beyond a certain distance from the nucleus, 
say $x_C$ (in the present calculation $x_C$ is set to 200 a.u. 
 (except for the results shown in Fig.~\ref{intrf}, see text),
an electron is unlikely to return.
Moreover, since $x_C$ is large, the Coulomb potentials 
are weak and the motion of the electron at that (or an even larger) distance is 
determined by the field only. 
Therefore, for distances larger than $x_C$, it can be assumed 
that the electron does not interact with the nucleus and with the other electron.
Such a description is reminiscent of the simple man's model \cite{corkum93} that
is frequently used to describe the rescattering scenario, but this analogy is only 
partially correct. In the simple man's model the interaction with the nucleus 
and the other electron is neglected directly after the tunnelling event, even when
the electron is still close to the nucleus. In our case, Coulomb interactions are 
neglected only for distances larger than $x_C$, where the assumption of a weak interaction
is more plausible.

The absence of Coulomb interactions simplifies the subsequent analysis considerably.
If one evolves the wavefunctionin the velocity gauge, 
the interaction with the laser field becomes multiplication by a phase in momentum space.
That in turn gives the possibility to efficiently evolve
a state in the momentum space as if the configuration space was infinite.
Specifically, at the beginning, when both electrons are ``close'' to the nucleus
(this region will be called $R_{in}$), i.e. $|r_i|<x_C$, the whole
evolution is described by the Hamiltonian corresponding to (\ref{ham})
but in the velocity gauge:
\begin{equation}
\label{ham_vg}
H_{in}=\sum_{i=1}^2\left(\frac{p_i^2}{2} +
\frac{\sqrt{3}}{2}A(t)p_i
-\frac{2}{|r_i|}\right)+
\frac{1}{\sqrt{(r_1-r_2)^2+r_1r_2}}.
\label{fullh}
\end{equation}
The wavefunction$\psi(r_1,r_2,t)$ spreads during the course of the evolution and
eventually some parts of it can enter the outside region $|r_i|>x_C$, though still remaining 
well within the extension $\lbrack -L/2,L/2\rbrack $ of the numerical grid. 
We want to project out the parts of the wavefunctionin this region and evolve them in a 
simplified way. This requires some care since  
the wavefunctionhas to be cut smoothly to minimize reflections on 
the boundaries. We know that this can be done using absorbing boundary conditions,
so we shall use a similar approach with the difference 
that the ``absorbed'' part of the wavefunctionwill be further evolved in a 
simplified way in the outer regions, as we discuss below \cite{distance}.

\begin{figure}[t]
\vspace{1.5cm}
\begin{center}
\includegraphics[width=0.4\textwidth,clip]{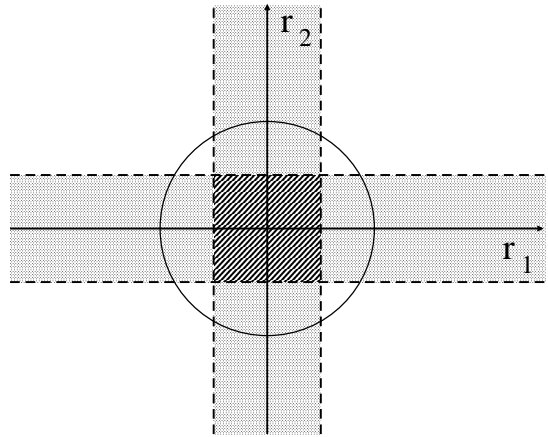}
\end{center}
\caption[]{Integration regions for the transfer of the wave functions 
between the different regions. 
Dashed lines correspond to borders between the full-integration region 
(dashed interior region, $R_{in}$), 
the full-simplified-integration region (white, $R_{out}$), and the 
semi-simplified-integration regions 
(shaded, $R_1$ and $R_2$). The circle indicates the area where the main 
density of the wavefunctionis located.
\label{ir}}
\end{figure}

To see how the transfer between inner and outer regions is calculated,
consider a wavefunction$\psi(r_1,r_2)$ that may extend (as indicated by the circle in 
Fig.~\ref{ir}) beyond the critical distance $x_C$. 
It can be written as a coherent sum of four parts, namely:
\begin{eqnarray}
\psi(r_1,r_2) &=& \psi_{in}(r_1,r_2) + \psi^1_{out}(r_1,r_2) \\ \nonumber
&&+ \psi^2_{out}(r_1,r_2) + \psi_{out}(r_1,r_2).
\end{eqnarray}

The $\psi_{in}$ part is the wavefunctionwith both electrons ``close'' 
to the nucleus, $|r_i|<x_C$: the wavefunctionis said to be in the $R_{in}$ region 
(dashed region in Fig.~\ref{ir}) (with exponential tails in the outside region). 
This part of the wavefunctionis evolved with the full Hamiltonian (\ref{fullh}). 

The $\psi^1_{out}$ part corresponds to the wavefunctionthat describes
a situation where the first electron has crossed
the border, but the other one not, i.e. $|r_1|>x_C$ and $|r_2|<x_C$
($R_1$ region, shaded area in Fig.~\ref{ir}).
Its evolution is governed by Hamiltonian
\begin{equation}
\label{ham_vg_1_simple}
H_1=\sum_{i=1}^2\left(\frac{p_i^2}{2} +
\frac{\sqrt{3}}{2}A(t)p_i\right)
-\frac{2}{|r_2|}.
\end{equation}
which neglects the Coulomb interaction of the outer (first) electron.
Thanks to such a simplification, the evolution of the first electron 
is just a multiplication by a phase factor in the momentum space. 
The evolution of the other electron still includes the interaction with 
the nucleus. 
The $\psi^2_{out}$ part has the same meaning, but with regard to the second electron. 
The last term, $\psi_{out}$, corresponds to the wavefunctionin the
region, $R_{out}$ (the white domains in Fig.~\ref{ir}), where Coulomb interactions are neglected. 
Integration here is done using the Hamiltonian
\begin{equation}
\label{ham_vg_2_simple}
H_{out}=\sum_{i=1}^2\left(\frac{p_i^2}{2} +\frac{\sqrt{3}}{2}A(t)p_i\right),
\end{equation}
for which it reduces to multiplication by appropriate phase factors in the momentum space.

In every time step the entire wavefunctionis evolved with the use of 
the above mentioned Hamiltonians. Then, parts of the wavefunctionthat cross
borders are cut and added to the wavefunctionin the appropriate regions.
 
First, consider the evolved function $\psi_{in}$ 
It is divided in two parts:
\begin{equation}
\psi_{in} = {\tilde \psi}_{in} + \Delta \psi_{in}.
\end{equation}
The first contribution acquires exponential-like tails outside of the inner region:
\begin{eqnarray}
{\tilde \psi}_{in} =\left\{
\begin{array}{ll}
\psi_{in}, & {\rm for} \quad|r_i|<x_C, \\
D(r_1)\psi_{in}, & {\rm for} \quad|r_1|>x_C, |r_2|<x_C, \\
D(r_2) \psi_{in}, & {\rm for} \quad|r_1|<x_C, |r_2|>x_C, \\
D(r_1)D(r_2)\psi_{in}, & {\rm for} \quad|r_i|>x_C, 
\end{array}\right.
\end{eqnarray}
where
\be
D(r_i)=\exp\left[-\eta(|r_i|-x_C)^\alpha\right].
\ee
Here $\eta$ and $\alpha$ are parameters chosen to minimize reflections 
and take the same values  as for the absorbing boundary conditions (see \ref{technical_numerics}). 
The part ${\tilde \psi}_{in}$ will become the new
$\psi_{in}$ function when the process of reorganizing the wavefunctionis completed.
The second term is just $\Delta \psi_{in}=\psi_{in} - {\tilde \psi}_{in}$ and it is decomposed 
into three terms:
\bea
\Delta\psi^1_{in}&=&\left\{
\begin{array}{ll}
\Delta\psi_{in}, & {\rm for} \quad |r_1|>x_C, |r_2|<x_C \\
D(r_2)\Delta\psi_{in}, & {\rm for} \quad |r_i|>x_C,
\end{array}
\right. \\
\Delta\psi^2_{in}&=&\left\{
\begin{array}{ll}
\Delta\psi_{in}, & {\rm for} \quad |r_1|<x_C, |r_2|>x_C \\
D(r_1)\Delta\psi_{in}, & {\rm for} \quad |r_i|>x_C,
\end{array}
\right. \\
\Delta\psi_{in}^{out}&=&\Delta \psi_{in}-\Delta\psi^1_{in}-\Delta\psi^2_{in}.
\eea
These terms will be coherently added to the wave functions in the appropriate regions.

Consider now the $\psi^1_{out}$ part evolved under (\ref{ham_vg_1_simple}) for
$|r_1|>x_C$.
The evolved function may extend also into the region where $|r_2|$ exceeds
$x_C$. Thus we define,
\bea
{\tilde \psi}^1_{out}=\left\{
\begin{array}{ll}
\psi^1_{out}, & {\rm for} \quad |r_2|<x_C, \\
D(r_2)\psi^1_{out}, & {\rm for} \quad |r_2|>x_C, 
\end{array}
\right.
\eea
and 
\be
\Delta\psi^1_{out}=\psi^1_{out}-{\tilde \psi}^1_{out}.
\ee
The part $\Delta \psi^1_{out}$ will be the contribution added to the ``double ionization'' 
sector later. The wavefunction$\psi^2_{out}$ is treated similarly, with $r_1$ and $r_2$
interchanged. If the wavefunctionhas a definite parity under exchange, this can
efficiently be implemented with the symmetrization \cite{eckhardt07}.

Assuming the trivial evolution for  $\psi_{out}$ has also been performed we
may update the wavefunction. 
The inner part is just replaced by ${\tilde \psi}_{in}$. 
The part that comes from the inner region is added to the wavefunction in the regions 1 (and 2) 
and, at the same time, the corresponding part that goes to the outer region is cut from it.
The wavefunctionin the outer region has only incoming contributions from all other regions. 
The update is, therefore, realized by the following sequence:
\begin{eqnarray}
\psi_{in} &\rightarrow& {\tilde \psi}_{in},\nonumber\\
\psi^1_{out} &\rightarrow& {\tilde \psi}^1_{out}+\Delta\psi^1_{in}\nonumber\\
\psi^2_{out} &\rightarrow& {\tilde \psi}^2_{out}+\Delta\psi^2_{in}\nonumber\\
\psi_{out} &\rightarrow& \psi_{out} + \Delta\psi^{out}_{in} +\Delta \psi^1_{out} + \Delta \psi^2_{out}.
\end{eqnarray}

This sequence is repeated in every time step and of course it is accompanied by
appropriate changes of representations from position space to momentum space.

Finally,  one can collect all parts of 
the wave function, add them coherently and 
obtain the wavefunction of the whole system in the momentum representation.

\subsection{Extracting two-electron information}

\begin{figure}[t]
\begin{center}
\includegraphics[height=0.3\textwidth,clip]{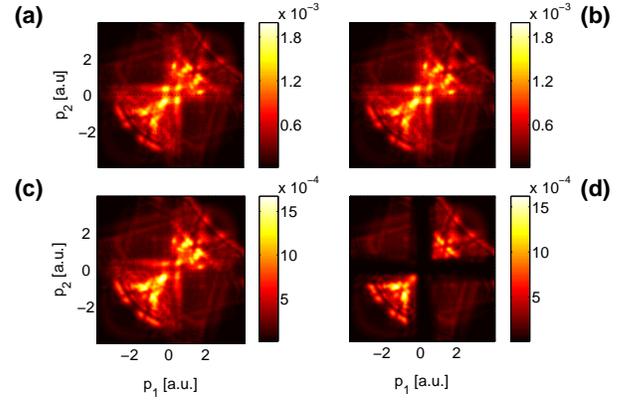}
\end{center}
\caption[]{(Color online) Effect of the ionization radius on the electron momentum
distribution: (a) $|r_i|>$10 a.u., (b) $|r_i|>$25 a.u., 
(c) $|r_i|>$50 a.u. and (d) $|r_i|>$200 a.u. 
The calculations are for a laser pulse with an amplitude $F_0=0.3$~a.u.,
a carrier-envelope phase $\phi=0$ and a duration of 5 cycles. 
The distributions are convoluted with a Gaussian of $\sigma_p=0.07$~a.u. 
width.
\label{rozne_paski}}
\end{figure}

We now have access to the whole momentum distribution of our system but
we have yet to give the prescription how to extract from the two electron wavefunction
the interesting information about the momenta of two ionized electrons. 
As has been already noted, one can define regions of the coordinate space that are identified 
with an atom, and singly or doubly charged ions. 
In order to focus on the double ionization, the parts of the wavefunction corresponding
to the atom and to the single ion may be smoothly cut out from the wave function. 
Those regions form a cross (compare again Fig.~\ref{geometry}). 
Its width may be altered and in such a way the size of the regions corresponding
to the atom, the single and the double ionization may be changed. 
That obviously affects the corresponding momenta distributions. 

In Fig.~\ref{rozne_paski}
momenta distributions for different cross widths are depicted. 
The last panel, (d), corresponds to the wave function
from the outer region, i.e. $R_{out}$ ---  $|r_i|>200$~a.u. 
As can be seen, by the cutting out the inner part of the wavefunction
identified with the atom or the singly charged ion, one erases small and uncorrelated
momenta from the distribution. 
That momenta are located along axes and close to the center of the coordinate system. 
Without much loss one could analyze only the part of the wavefunctionthat lies
 in the $R_{out}$ region -- see  Fig~\ref{rozne_paski}(d) -- 
because it already reveals the key feature of the 
correlated escape, i.e. a significant population along the diagonal, $p_1=p_2$.
However, comparison with the panel (c) shows
that one neglects some of electrons moving in a correlated manner with smaller momenta
(note the maxima along the diagonal close to the center of the coordinate system), 
that do not reach the $R_{out}$ region before the end of integration. 
 Therefore, we have decided that in the following analysis momentum distributions corresponding to
the case of $|r_i|>50$~a.u. will be considered.

\begin{figure}[t]
\begin{center}
\includegraphics[height=0.3\textwidth,clip]{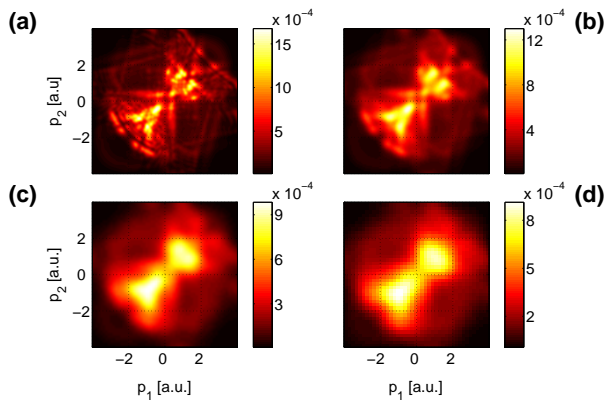}
\end{center}
\caption[]{(Color online) Resolution dependence of the final momentum distributions:  
the distributions are smoothed with Gaussians of width (a) 0.07 a.u., (b) 0.16 a.u., 
(c) 0.3 a.u., and (d) 0.4 a.u, respectively. 
The calculations are for a pulse with 
an amplitude $F_0=0.3$~a.u., a carrier-envelope phase $\phi=0$ and 5 cycles duration. 
Only the wavefunctionin the region $|r_i|>50$~a.u. is included in the calculation.
\label{rozne_rozdz}}
\end{figure}

Next we examine the variations of the distributions with resolution.
In numerical simulations one could resolve the distributions arbitrarily well, 
however, at the expense of very long integration times and increasing memory requirements.
On the other hand, the experimental distributions are obtained with a finite resolution, only.
In order to compare the data obtained with the present model with the experimental ones, 
we convolute all momenta distributions with Gaussians.
To see the effect of this smoothing, momentum distributions obtained with different resolutions,
from the presently best experimental resolution of $\sigma_p=0.07$ a.u.~\cite{weckenbrock04} up
to a value of $\sigma_p=0.4$, are compared in Fig.~\ref{rozne_rozdz}. 
Panels (c) and (d) resemble experimental distributions~\cite{weber00n}
very much, even though only one carrier-envelope phase was used. 
Panel (d) corresponds to the experimental resolution
obtained in the first experiment in which the momenta distributions 
were measured~\cite{weber00n}. 
Panels (a) and (b) show considerable substructure, which may have its origin in quantum interferences
(see below).

\section{Results for smooth pulses}
\label{pulses}

A first batch of results obtained using the above prescription were presented
in \cite{prauzner07}. 
There the focus was on flat-top (trapezoidal shaped) pulses that allowed for a 
clear identification of the relation between the pulse intensity (being constant for the most of the 
pulse duration) and the ionization dynamics. In another short contribution \cite{eckhardt07} 
we have concentrated on the influence of the symmetry of the wavefunctionon the correlated electron 
escape. Here, apart from providing some of the technical details, 
we shall concentrate 
on smooth laser pulses with electric field amplitude of the form (\ref{envelope}). In all the calculations 
we fix the frequency of the field to be $\omega=0.06$ a.u.

\begin{figure}[t]
%\vspace{0.5cm}
\begin{center}
\includegraphics[width=0.4\textwidth,clip]{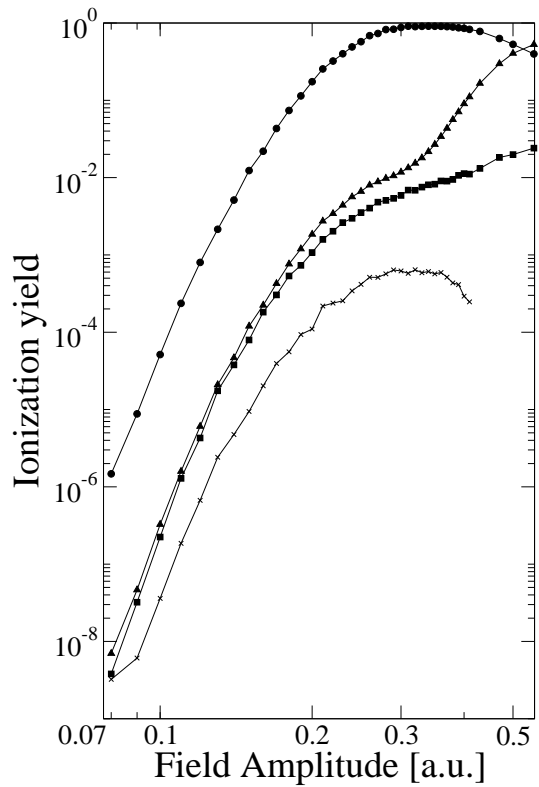}
\end{center}
\caption[]
{Ionization yields versus field strength. From top to bottom: single ionization (circles), non-simultaneous 
double ionization (triangles), simultaneous double ionization (squares) and anti-correlated double 
ionization (crosses). The calculations are for a pulse of five cycles duration.
}
\label{yield}
\end{figure}

\begin{figure}[t]
\begin{center}
\includegraphics[width=0.4\textwidth,clip]{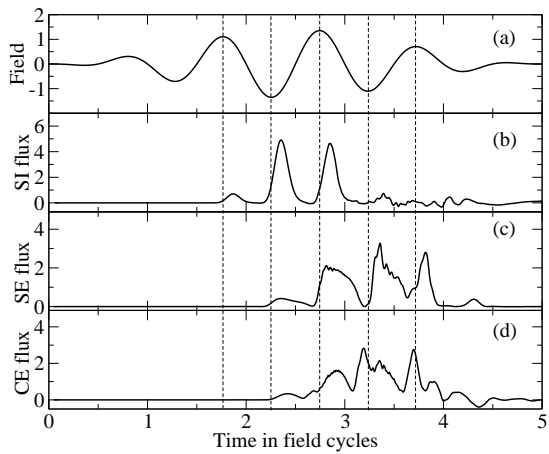}
\end{center}
\caption[Time dependence of probability fluxes for single, double sequential 
and non-sequential ionization for 5 cycle pulse]
{Probability fluxes  as a function of time. Panel (a) represents the instantaneous field strengths for 
a pulse with maximal field amplitude $F_0=0.18$ and 5 cycle duration. 
The flues related to single ionization, to simultaneous ejection and to other
consecutive
double ionization processes are shown in panels (b), (c) and (d), respectively.
All fluxes are in arbitrary units (in particular, recall from Fig.~\ref{yield} that single ionization 
fluxes are typically hundreds of times higher than double ionization fluxes).}
\label{time}
\end{figure}

Consider first the  total ionization yields, shown in Fig.~\ref{yield} for a pulse of five 
cycles duration and a fixed phase $\phi=0$. The single ionization yield shows the typical behaviour, it 
reaches almost 100\% for  high field intensities, and then drops down for still higher fields when 
the double ionization becomes significant. 
We observe that: 
\begin{enumerate}
\item up to a field strength $F_0\approx 0.3$, the double ionization yields are several orders
of magnitude below the single ionization yield, 

\item  simultaneous escapes increase strongly until about
$F_0=0.2$, where the increase becomes noticeably slower,
\item consecutive escapes pick up near about $F_0=0.3$, thereby accounting almost
exclusively for the knee structure familiar from experimental observations,
\item  for five cycle pulses considered here, SE is always 
less probable than CE  where one electron ionizes after the other; 
the difference
is small for low field amplitudes and increases for higher field amplitudes,
\item the signal for anti-correlated electron escape 
is several orders of magnitude smaller than that for other processes. Its non-zero value may actually
have a numerical origin, and be connected with the calculation of the flux at a 
finite distance from the nucleus.
\end{enumerate}
The  consecutive ionisation process may consist of at least two paths: the independent electron escape where, 
after the escape of the first electron, the other has to get rid of the nucleus' attraction on 
its own (this process becomes more significant for $F_0$ above the knee) and the process where due to 
the rescattering the second electron is first excited and then, after the first electron is already
gone, it leaves the ion too. The rescattering may be repeated several times, thus contributing to an increase 
of both the simultaneous and consecutive escapes. 

More information about the dynamics of the ionization process, in particular about
the temporal sequence of events and the timing of double ionization during the
pulse, may be extracted by following the time dependencies of the probability fluxes. 
They are shown in Fig.~\ref{time}, together with the field amplitude.
Note the strong correlation between panels (b) and (c) with pronounced
maxima for the SE flux occurring about half a cycle after the SI flux maxima:
this is a direct confirmation of the  rescattering scenario. 
First, the electron leaves the {\bf A} region, evidenced  
in Fig.~\ref{time}b by the peaks in the single ionization flux near the field extrema, 
when the saddle for the single ionization is open and the ionization itself is the most probable.
Then the electron is turned back to the nucleus and gains energy as it is accelerated by the field. 
It hits its parent ion and shares its energy with the electron near the core to form a
highly excited state. This highly
excited compound state then decays through single, 
simultaneous escape and 
consecutive double ionization. 
 This is seen as peaks in fluxes into both channels
that appear roughly one half-cycle later than the corresponding peaks in the single 
ionization flux (see Fig.~\ref{time}).
In the first half of the pulse single ionization dominates; then, once there is enough 
time for the electron to turn back and to rescatter, double ionization follows, see 
Fig.~\ref{time}c and d.
Moreover, both types of double ionization events occur close to the laser field extrema. 
In the case of simultaneous double escape, this means that the field is sufficiently 
large to open the saddle for the symmetric escape. This observation is at variance
with expectations based on the simple man model \cite{corkum93}, but in agreement with 
previous deductions from the analysis of the classical paths \cite{eckhardt01pra1,eckhardt06}.

\begin{figure}
\begin{center}
\includegraphics[width=0.4\textwidth,clip]{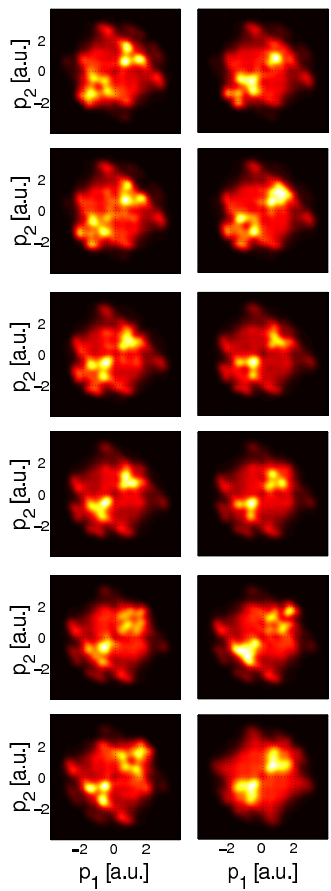}
\end{center}
\caption[]{(Color online) Phase dependence of the electron momentum distributions. The phases increase from left to
right and top to bottom from $\phi=0$, $0.1\pi$, $...$, to $\pi$. Note that the frame on the
bottom left for $\phi=\pi$ is the mirror image of the one on the top left for $\phi=0$, 
reflecting the mirror symmetry in the pulses. The frame on the bottom right is obtained
for a uniform average over all phases. The calculations are for a pulse with 
$F_0=0.2$ and a five cycle duration. The distributions are convoluted with 
a Gaussian of $\sigma_p=0.2$~a.u. width, and only the wave functions in the 
region $|r_i|>50$~a.u. is shown. 
\label{mome1}}
\end{figure}

The electron momenta distributions for different carrier-envelope phases are shown in Fig.~\ref{mome1}. 
The chosen value of the field amplitude $F_0$ corresponds to the beginning of the knee structure in 
Fig.~\ref{yield}. Observe that, regardless of the phase, a significant fraction of the electrons are ejected 
with similar momenta $p_1\approx p_2$, as the distributions are diagonally dominated.

The momentum distributions show a significant dependence on the carrier envelope phase of the pulse. 
Apart from the obvious global symmetry corresponding to the change of 
$\phi$ by $\pi$, which is equivalent to the change of the sign of the momenta, 
one observes rapid changes of distributions corresponding to small changes of phases 
(compare e.g.  panels corresponding to $\phi=0.1\pi$ and $\phi=0.2\pi$, or  $\phi=0.5\pi$ and $\phi=0.6\pi$). 
This behaviour conforms with classical and analytical studies \cite{liu04,maciek2004},
that even propose to use the momentum distributions for the identification of the 
laser pulse phase. We will return to the differences in fine-structure in the next section.

\begin{figure}
\begin{center}
\includegraphics[width=0.4\textwidth,clip]{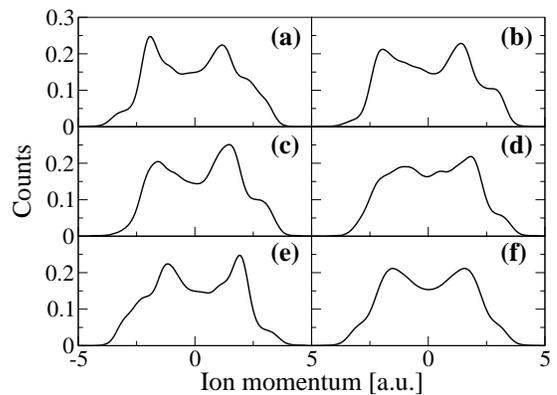}
\end{center}
\caption[Ion Momenta distribution for different phases]
{Phase difference of the ion momentum distribution. The phases are 
(a) $\phi=0$, (b) $3\pi/10$, (c) $\pi/2$, (d) $4\pi/5$, and (e) $\pi$.
Note that the distributions for $\phi=0$ and $\phi=\pi$ in (a) and (e) differ
by a reflection of the ion momentum, only.
Panel (f)  shows the distribution averaged over a uniform $\phi$ distribution. 
The calculations are for a pulse with $F_0=0.2$ and five cycle 
duration. 
} 
\label{iondis}
\end{figure}

In addition to the electron momentum distributions, we can also study the ion momentum distribution,
see Fig.~\ref{iondis}. Within the model, the ion distribution is nothing but the negative of the
sum of electron momenta, corrected by the geometrical factor due to the $\pi/3$ angle between the axes,
(see Fig.~\ref{geo1}), i.e. $p_{\rm ion}=-\sqrt{3}(p_1+p_2)/2$. When averaged over 
a uniform distribution of phases under a carrier envelope this distribution is symmetric and shows 
the celebrated double-hump structure. 

\section{Interference fringes in momenta distributions}
\label{fringes}

In almost all momenta distributions presented until now one could notice substructures that are reminiscent
of quantum interference patterns. Within a semi classical picture, the coherence of the wavefunctionis
maintained, and the pattern could arise from interferences between contributions from several
paths to the same final momenta. Some of these paths have been seen within classical trajectory studies
\cite{eckhardt06}. Arb\'o {\it et al.} \cite{burgdoerfer} have
also shown that ionization events
originating at different times during a pulse can contribute to the ionization signal and give
rise to interferences. In all cases the substructures require a good resolution in order
to be resolved (recall the effects of resolution shown in Fig.~\ref{rozne_rozdz}). Therefore, 
in this section, we shall work with a resolution of $0.07$~a.u., as is accessible in the best 
current experiments \cite{weckenbrock04}.

\begin{figure}[ht]
\begin{center}
\includegraphics[height=0.3\textwidth,clip]{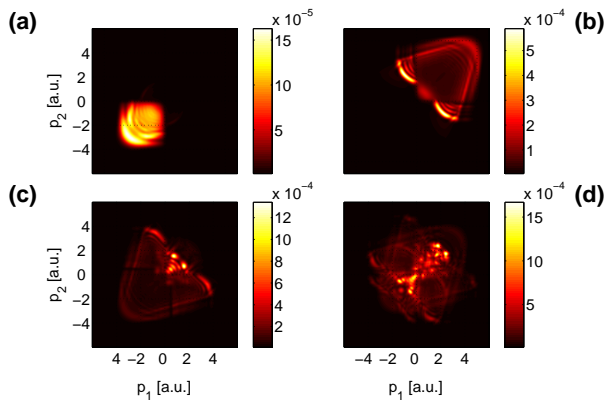}
\end{center}
\caption[]{(Color online) Effect of pulse length on momentum distributions. 
The length of the pulse is (a) one cycle, (b) two cycles, (c) three
cycles, (d) four cycles. 
The calculations are for a pulse with an amplitude $F_0=0.3$ a.u., and
a carrier-envelope phase $\phi=0$. 
The wavefunctioncorresponding to the region $|r_i|>50$~a.u. is shown.
The distributions are convoluted with a Gaussian of width $\sigma_p=0.07$~a.u. 
\label{intrd}}
\end{figure}

\begin{figure}[ht]
\begin{center}
\includegraphics[width=0.4\textwidth,clip]{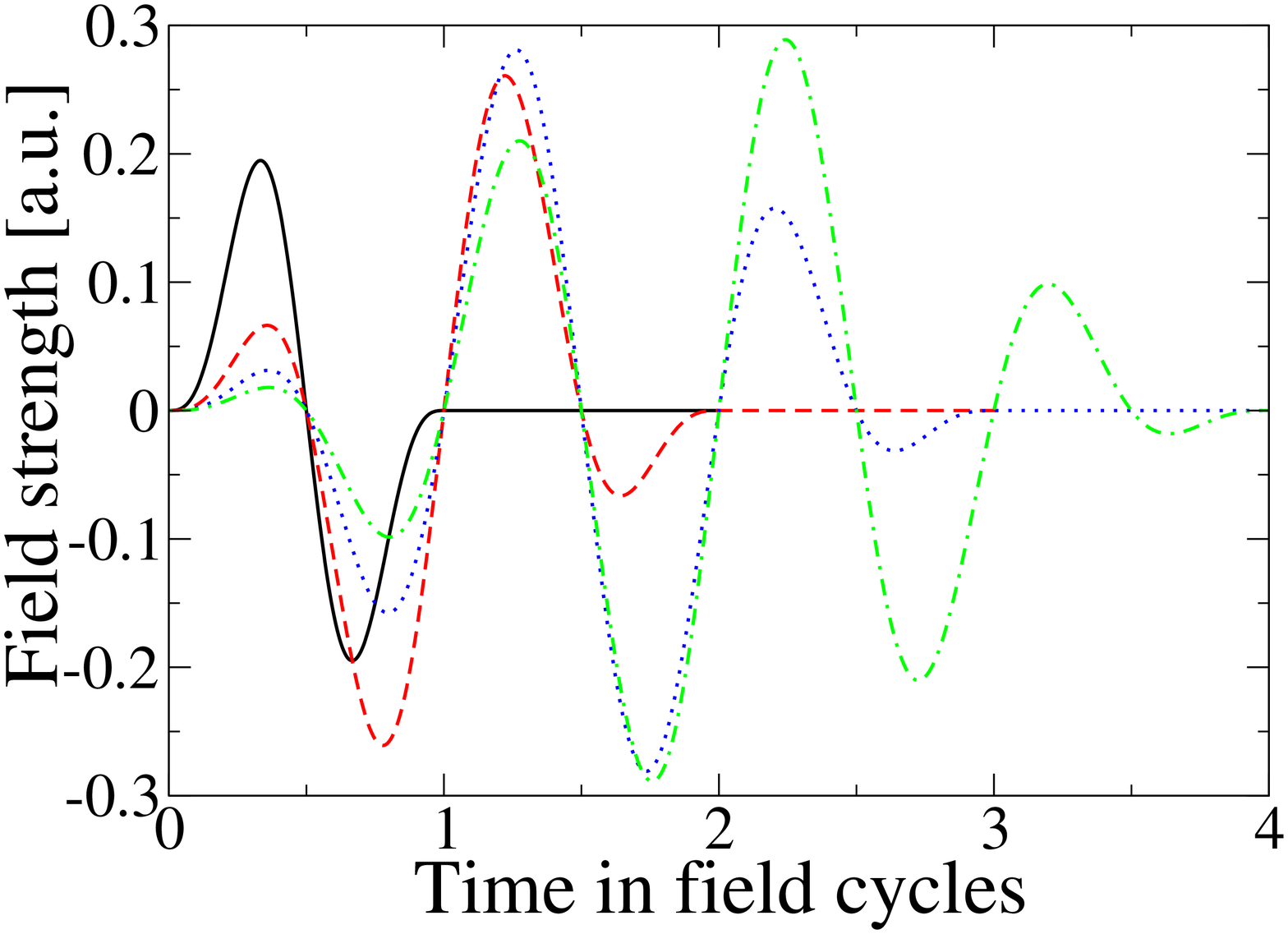}
\includegraphics[width=0.4\textwidth,clip]{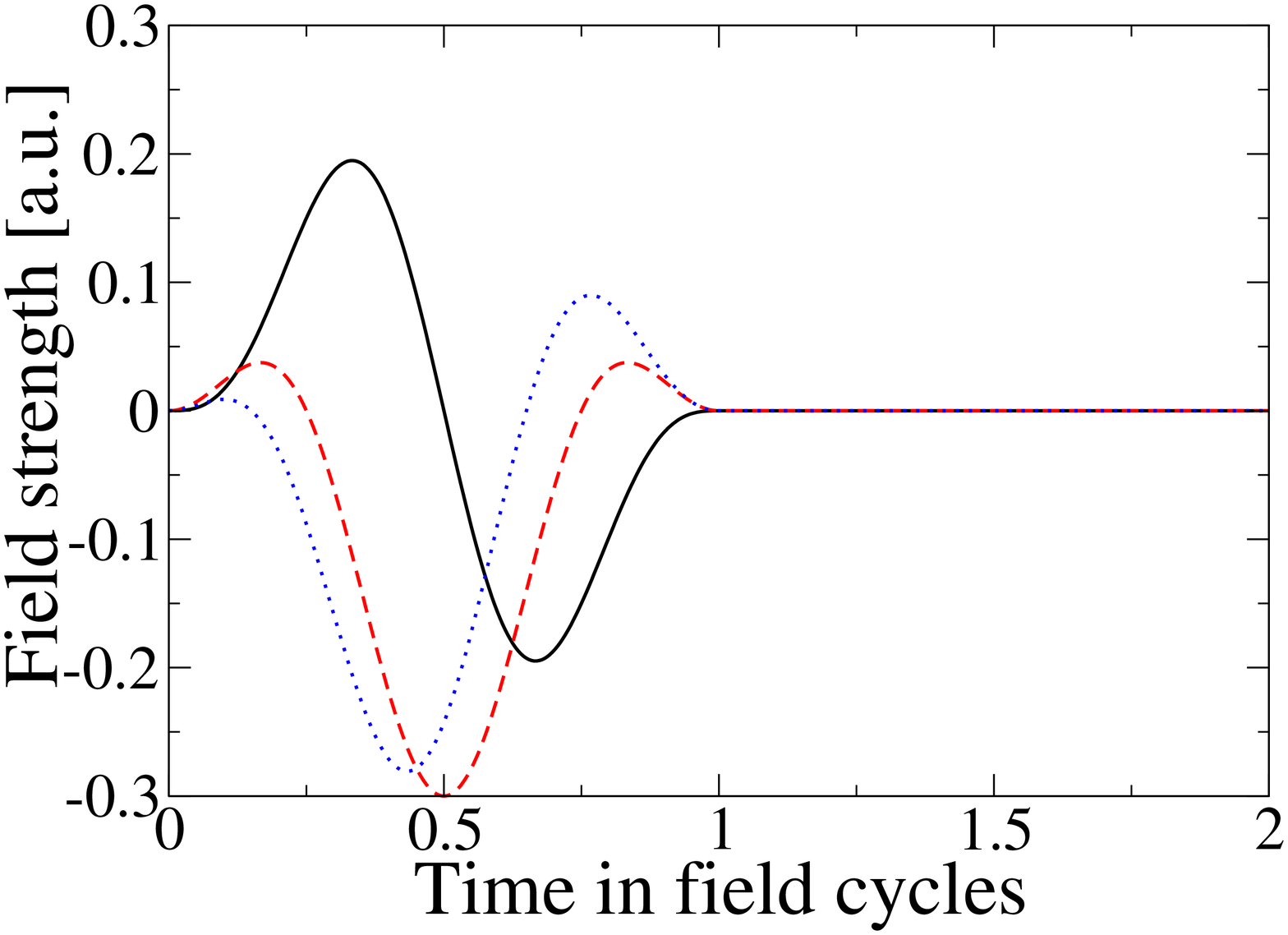}
\end{center}
\caption[]{(Color online) Variations in instantaneous field strength with duration and phase of the pulse. 
The pulses have  a field amplitude
$F_0=0.3$ a.u.. For the top panel, the carrier-envelope phase is set to $0$ , and the
duration is one cycle  (black solid line), two cycles (red broken line), three cycles (blue dotted line) and
four cycles (green dash-dotted line). For the bottom panel, the duration of the pulse is fixed to
be a single cycle, and the phases are $0$ (black solid line), $0.5\pi$ (red broken line) 
and $0.7\pi$ (blue dotted line).}
\label{pl}
\end{figure}

The observed pattern depends on several parameters of the problem, including
the amplitude of the field,  the pulse duration and the carrier-envelope phase. 
Among them, the pulse duration seems to be the most critical for a theoretical understanding
of the process, in that the shorter the pulse, the smaller the number of possible paths that
lead to double ionization.
The increase in complexity is demonstrated in Fig.~\ref{intrd}, where momenta distributions 
for pulses of different duration are shown. 
Apparently, each additional cycle increases the complexity of the pattern.

However, the pulse duration is not the only reason for the observed differences.
 Because of the ramping of the field, the actual maximal instantaneous field strength 
that is reached during the pulse may be lower than the nominal amplitude $F_0$, 
and may vary with the envelope function and the phase of the field, see Fig.~\ref{pl},
where the actual field amplitudes for the momentum distributions of Fig.~\ref{intrd}
are shown. The  black solid line in the top panel
 corresponds to the single cycle pulse and the maximal instantaneous 
effective field amplitude is $F_{max}=0.19$ a.u., even though the nominal pulse amplitude 
is set to be $F_0=0.3$ a.u. For the longer pulses the maximal field
amplitude becomes $F_{max}=0.25$, $0.28$, and $0.29$ a.u. for pulses with
two, three and four cycles, respectively. The maximal field strength agrees with $F_0$ for longer pulses.

\begin{figure}[ht]
\begin{center}
\includegraphics[width=0.45\textwidth,clip]{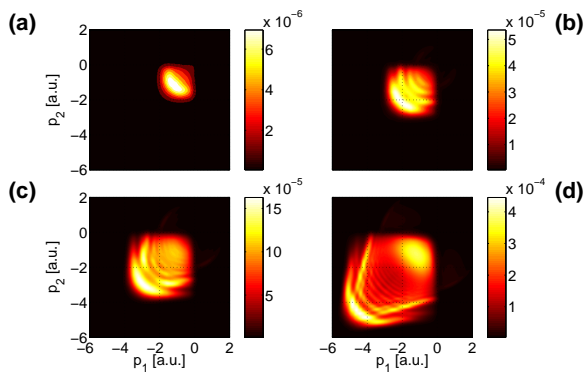}
\end{center}
\caption{{(Color online)  Electron momentum distributions corresponding 
to the single cycle laser pulse 
for $\phi=0$ and $F_0=0.2$, $0.25$, $0.3$, and $0.4$  [from (a) to (d)]. 
Note that the interference fringes appear for larger field amplitude only.
The wavefunctioncorresponding to the region  $|r_i|>50$~a.u. is shown.
The distributions are convoluted with a Gaussian of $\sigma_p=0.07$~a.u. width. 
\label{s2}}}
\end{figure}

To simplify the picture we shall consider in the following the single cycle pulse only. 
Fig.~\ref{s2} shows the momentum distributions for a sine-squared envelope and
different field amplitudes $F_0$. 
For higher field amplitudes $F_0$, the momentum distribution reveals interesting
interference patterns. While the detailed analysis of their origin requires extensive semiclassical studies that are beyond the scope of the present paper,
we attempt below a qualitative understanding of the phenomenon.

Let us now consider how the absolute phase of the laser pulse affects the interference pattern. This dependence can be expected to be quite dramatic,
in view of earlier observations for  few cycle pulses (compare 
Fig.~\ref{mome1}). Note also the strong dependence of the effective electric 
field temporal behaviour on the phase, Fig.~\ref{pl}. It turns out that
 the interference fringes are present  
for the carrier-envelope phase close to $\phi=0$ only. Then the absolute value of the field has two maxima of comparable amplitude. For other values of the phase
a single maximum (of the absolute value) dominates the field temporal behaviour.

That in turn points out strongly towards a rescattering scenario as a necessary ingredient of the appearance of the fringes. The first field maximum excites a single electron, it turns back after the field direction changes and as a highly
energetic particle shares its energy with the remaining electron leading to a 
double ionization. Such a process is less probable for a singly peaked field.

\begin{figure}[t]
\begin{center}
\includegraphics[width=0.45\textwidth,clip]{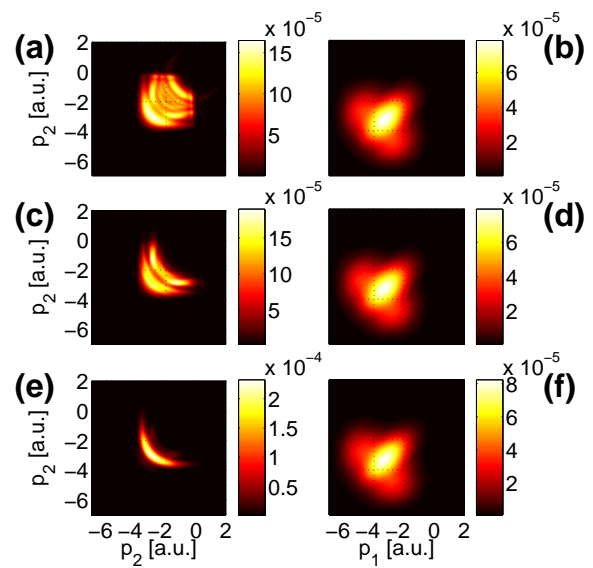}
\end{center}
\caption[]{(Color online) Effect of rescattering on momentum distributions. (a) and (b) show
the momentum distributions for $\phi=0$ and $\phi=0.5\pi$, respectively,
as obtained with the cut-off $x_C=50$~a.u.. For (c) and (d), the cut-off
is reduced to $x_C=25$~a.u., and for (e) and (f) to $x_C=12.5$~a.u..
Decreasing the cut-off reduces the
contributions from rescattered electrons, and the contrast in the interference pattern.
The calculations are for a laser pulse with  an amplitude $F_0=0.3$ a.u. 
and a pulse duration of one cycle. 
The momentum distributions are calculated from wave functions in the regions 
$|r_i|>50$~a.u. and are convoluted with a Gaussian of width $\sigma_p=0.07$~a.u. 
\label{intrf}}
\end{figure}

In order to study further the significance of the rescattering for the
presence of the fringes, we repeated the time evolution with
different (smaller)  $x_C$ which determines the ionization regions,  
thereby limiting the electronic motion (leading to rescattering) to
area close to nucleus (once an electron reaches the distance $x_C$ from 
the nucleus its evolution is simplified by 
neglecting Coulomb potentials, see Sec.~IIID). Such an approach has been
used recently to discuss the influence of long rescattering orbits in
non sequential double ionization of the hydrogen molecule \cite{baier06}.
We focus   on momentum distributions 
for two phases; those for $\phi=0$ that show interference patterns, and the other for $\phi=0.5\pi$ that do not show them in Fig.~\ref{intrf}.  
The gradual disappearance of the 
fringes in the left column when $x_C$ decreases suggests that the existence of the interference pattern is linked to the rescattering process.

The situation is not that simple, however. Additional classical trajectory studies indicate that even  for small $x_C=12.5$~a.u. values there exist rescattering trajectories; those, however, extend less from the nucleus. The fringes are
thus most probably due to the quantum interference of contributions coming
from ``long'' and ``short'' rescattering orbits. The latter only are eliminated by
restricting the interaction between the electrons to the area very close to the
nucleus. 

The interference phenomena due to ``long'' and ``short'' trajectories has been
discussed in the context of high-harmonic generation several years ago 
\cite{lewenstein94,becker94}. It seems that the very same trajectories are responsible for the interference pattern in the momenta distribution of the
outgoing electrons in double ionization process.
 
The absence of any changes in the right column suggests
that for this phase of the field the contribution of long rescattering orbits is smaller,
hence supporting a more direct  double ionization process.  Note that in the latter case, i.e. 
$\phi=0.5\pi$, the field possesses very strong maximum in the middle of the pulse (see
Fig.~\ref{pl}b) that weakens the rescattering effect.

\begin{figure}[t]
\begin{center}
\includegraphics[width=0.4\textwidth,clip]{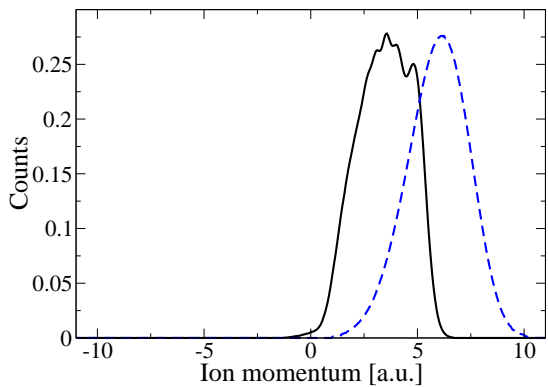}
\end{center}
\caption[]{
(Color online) Ion recoil momenta distributions corresponding to electron momenta distributions of 
Fig.~\ref{intrf}. Black solid line corresponds to
 $\phi=0$ and reveals a ragged top with several maxima, reflecting the 
interferences shown in left panels of  Fig.~\ref{intrf}.
The  dashed line represents the distribution  for $\phi=0.5\pi$; it
is smooth, in accord with the absence
of interferences in the right column of Fig.~\ref{intrf}
 \label{iond3}}
\end{figure}

As discussed before, the electron momentum
distributions yield in a direct way the ion recoil momentum distributions.
The latter, for the field strengths and phases of Fig.~\ref{intrf}, 
are presented in 
Fig.~\ref{iond3}. The presence of interferences in the electron momenta shows up
in the ion momentum distribution in the form of a ragged maximum. 
Since ion recoil momentum distributions can be measured experimentally, 
the transition between ragged and smooth maxima should be readily accessible to experimental
verification.

\section{Conclusions}
\label{end}
The calculation of wave functions and ionization events and their interpretation
presented here show that the reduced dimensionality model proposed
in \cite{eckhardt06} captures many of the features of the quantum
double ionization process. The knee structure in the ionization yield, 
the double hump structure in the ion momentum distribution and the electron momentum 
distributions are in agreement with observations, and the results on the phase
dependence and the interference patterns suggest further experimental and theoretical
studies. The major advantage of the present 
model over the well known and much studied aligned
electron model is that it does allow for the correlated two-electron escape which is
blocked in the aligned model by the overestimated Coulomb repulsion. 

We have discussed the details of the numerical implementation of the model as well as
the methods that enable us to extract the relevant physical information from the
dynamically evolved wavefunction. We have discussed the ionization features for 
smooth few cycle pulses as well as for very short pulses 
(up to a single cycle pulses). For the former we have 
confirmed that the model yields prediction in a qualitative agreement with experimental 
data for all the calculated observables (ion yield, ion recoil and electron 
momenta distributions). We have confirmed previous suggestions 
\cite{liu04,maciek2004} that the absolute phase of the
pulse carrier envelope may be extracted from the momenta distributions. 
For even shorter pulses the phase dependence of the signals becomes quite dramatic. 
The electron momentum distributions show distinct interference patterns which 
are traces of different paths leading to double ionization. With a sufficient 
resolution these interference patterns can also be seen in longer pulses, 
however their complexity grows significantly with the pulse duration. 
The fringes provide additional information about the absolute phase  but also 
are manifestations of nontrivial dynamics. 
Interestingly, their very existence seems 
to be intimately connected to the rescattering process. 
The interferences can also be seen in ion recoil momentum distributions, and hence
should be quite easily accessible in experiments that do not resolve the electron momentum
distributions.

\section{Acknowledgements}

A significant part of the numerical simulations were done at ICM UW under grant G29-10.
The work has been supported by the Deutsche Forschungsgemeinschaft and Marie Curie
ToK project COCOS (MTKD-CT-2004-517186). Support by the Polish Government 
scientific funds (2005-2008) as a research project is acknowledged.


\begin{thebibliography}{10}

\bibitem{gelt88} S. Geltman and J. Zakrzewski,  J. Phys.  B {\bf 21}, 47 (1988).

\bibitem{anne82} A. L'Huillier, L. A. Lompre, G. Mainfray, and C. Manus, 
Phys. Rev. Lett. {\bf 48}, 1814 (1982).

\bibitem{review} A. Becker, R. D\"orner, and R. Moshammer, 
J. Phys. B {\bf 38}, S753 (2005).

\bibitem{walker94} B. Walker {\it et~al.}, Phys. Rev. Lett. {\bf 73}, 1227 (1994)

\bibitem{weber00n}
T. Weber {\it et~al.}, Nature {\bf 405},  658  (2000);
T. Weber {\it et~al.}, Phys. Rev. Lett. {\bf 84},  443  (2000);
R. Moshammer {\it et~al.}, Phys. Rev. Lett. {\bf 84}, 447 (2000).
%R. Moshammer {\it et~al.}, J. Phys. B {\bf 36}, L113 (2003).

\bibitem{becker00kopold00}
A. Becker and F.~H.~M. Faisal, Phys. Rev. Lett. {\bf 84},  3546  (2000);
R. Kopold {\it et~al.}, Phys. Rev. Lett. {\bf 85}, 3781  (2000).

\bibitem{aligned}
R. Grobe and J. H. Eberly, Phys. Rev. Lett. {\bf 68}, 2905 (1992);
R. Grobe and J. H. Eberly, Phys. Rev. A {\bf 48}, 4664 (1993);
D. Bauer, Phys. Rev. A {\bf 56}, 3028 (1997);
D. G. Lappas and R. van Leeuwen, J. Phys. B {\bf 31}, L249 (1998);
W.-C. Liu,  J.~H. Eberly, S.~L. Hann, and R. Grobe, Phys. Rev. Lett. {\bf 83}, 521  (1999); 
S.~L. Haan {\it et~al.}, Phys. Rev. A {\bf 66}, 061402(R)  (2002).

\bibitem{engel}
M. Lein, E. K. U. Gross, and V. Engel, Phys. Rev. Lett. {\bf 85}, 4707 (2000).

\bibitem{becker}
C. Ruiz {\it et~al.}, Phys. Rev. Lett. {\bf 96}, 053001 (2006).  

\bibitem{taylor}
D. Dundas, K.~T. Taylor, J.~S. Parker, and E.~S. Smyth, 
J. Phys. B {\bf 32},  L231  (1999);

\bibitem{parker06}
J. Parker {\it et~al.},  
%B. J. S. Doherty, K. T. Taylor, K. D. Schultz, C. I. Blaga, and L. F. DiMauro, 
Phys. Rev. Lett. {\bf 96}, 133001 (2006).

\bibitem{corkum93}
P.~B. Corkum, Phys. Rev. Lett. {\bf 71},  1994  (1993);
K. Kulander, J. Cooper, and K. Schafer, 
Phys. Rev. A {\bf 51},  561  (1995).

\bibitem{eckhardt01pra1}
K. Sacha and B. Eckhardt, Phys. Rev. A {\bf 63},  043414  (2001);
B. Eckhardt and K. Sacha, Europhys. Lett. {\bf 56}, 651 (2001).

\bibitem{eckhardt06}
B. Eckhardt and K. Sacha, J. Phys. B: At. Mol. Phys. {\bf 39}, 3865 (2006).

\bibitem{prauzner07}
J. S. Prauzner-Bechcicki, K. Sacha, B. Eckhardt, and J. Zakrzewski,
Phys. Rev. Lett. {\bf 98}, 203002 (2007).

\bibitem{jensen84} R. V. Jensen, Phys. Rev. A{\bf 30}, 386 (1984).

\bibitem{leopold89} 
J.G. Leopold and D. Richards, J. Phys B{\bf 22}, 1931 (1989); {\it ibid.} {\bf 24}, 1209 (1991).

\bibitem{eberylium}
Q. Su, J. H. Eberly, and J. Javanainen, Phys. Rev. Lett {\bf 64}, 862 (1990);
V. C. Reed, P. L. Knight, and K. Burnett, Phys. Rev. Lett {\bf 67}, 1415 (1991).

\bibitem{Ho2005} P.J. Ho and J.H. Eberly, Optics Express {\bf 11}, 2826 (2003);
P.J. Ho, R. Panfili, S. L. Haan, and J.H. Eberly, Phys. Rev. Lett., {\bf 94}, 093002 (2005);
P.J. Ho and J.H. Eberly, Phys. Rev. Lett., {\bf 95}, 193002 (2005).

\bibitem{RESI}
B. Feuerstein {\it et~al.}, Phys. Rev. Lett. {\bf 87}, 043003 (2001);
V.L.B. de Jesus {\it et~al.}, J. Phys. B. {\bf 37}, L161 (2004).

\bibitem{footnote1}It is worth stressing that the probability current 
has this form only in the position gauge. In the velocity gauge it would also 
depend on the vector potential, $A(t)$.

\bibitem{distance} In fact the distance $x_C$, at which we neglect the Coulomb interactions, and $x_0$, at which the boundary conditions start to act, are 
taken to be the same in the numerical implementation.

\bibitem{eckhardt07} B. Eckhardt, J. S. Prauzner-Bechcicki, K. Sacha, and J. Zakrzewski, 
Phys. Rev. A {\bf 77}, 015402 (2008).% in press, see  arXiv:0705.3315 (2007).

\bibitem{liu04} X. Liu and C. Figueira de Morrison Faria,
 Phys. Rev. Lett. {\bf 92}, 133006 (2004).
 
\bibitem{maciek2004}C. Figueira de Morrison Faria, X. Liu, A. Sanpera, and 
M. Lewenstein, Phys. Rev. A {\bf 70}, 043406 (2004).

\bibitem{burgdoerfer}
D.G. Arb\'o, S. Yoshida, E. Persson, K. I. Dimitriou, and J. Burgd\"orfer, 
Phys. Rev. Lett. {\bf 96}, 143003 (2006); 
D.G. Arb\'o, E. Persson, and J. Burgd\"orfer, 
Phys. Rev. A {\bf 74}, 063407 (2006).

\bibitem{baier06} S. Baier, C. Ruiz, L. Plaja, and A. Becker,
Phys. Rev. A {\bf 74}, 033405 (2006).

\bibitem{lewenstein94} M. Lewenstein et al., Phys. Rev. A {\bf 49}, 2117 (1994).

\bibitem{becker94} W. Becker, S. Long, and J.K. McIver, 
 Phys. Rev. A {\bf 50}, 1540 (1994).
 
\bibitem{weckenbrock04}
M. Weckenbrock {\it et al.}, Phys. Rev. Lett. {\bf 91}, 123004 (2003);
M. Weckenbrock {\it et al.}, Phys. Rev. Lett. {\bf 92}, 213002 (2004).


\end{thebibliography}
\end{document}